\documentclass[iicol,sn-mathphys-num]{sn-jnl}

\usepackage{graphicx}%
\usepackage{multirow}%
\usepackage{amsmath,amssymb,amsfonts}%
\usepackage{amsthm}%
\usepackage{mathrsfs}%
\usepackage[title]{appendix}%
\usepackage{xcolor}%
\usepackage{textcomp}%
\usepackage{manyfoot}%
\usepackage{booktabs}%
\usepackage{algorithm}%
\usepackage{algorithmicx}%
\usepackage{algpseudocode}%
\usepackage{listings}%

\theoremstyle{thmstyleone}%
\theoremstyle{thmstyletwo}%

\theoremstyle{thmstylethree}%

\begin{document}

\title[Sociophysics models inspired by the Ising model]{Sociophysics models inspired by the Ising model}


\author*[1]{\fnm{Pratik} \sur{Mullick}}\email{pratik.mullick@pwr.edu.pl}

\author[2]{\fnm{Parongama} \sur{Sen}}\email{parongama@gmail.com}


\affil*[1]{\orgdiv{Department of Operations Research and Business Intelligence}, \orgname{Wrocław University of Science and Technology}, \orgaddress{\street{Wybrzeze Stanislawa Wyspianskiego 27}, \city{Wrocław}, \postcode{50-370}, \country{Poland}}}

\affil[2]{\orgdiv{Department of Physics}, \orgname{University of Calcutta}, \orgaddress{\street{92 APC Road}, \city{Kolkata}, \postcode{700009},  \country{India}}}




\abstract{The Ising model, originally developed for understanding magnetic phase transitions, has become a cornerstone in the study of collective phenomena across diverse disciplines. In this review, we explore how Ising and Ising-like models have been successfully adapted to sociophysical systems, where binary-state agents mimic human decisions or opinions. By focusing on key areas such as opinion dynamics, financial markets, social segregation, game theory, language evolution, and epidemic spreading, we demonstrate how the models describing these phenomena, inspired by the Ising model,  capture essential features of collective behavior, including phase transitions, consensus formation, criticality, and metastability. In particular, we emphasize the role of the dynamical rules of evolution in the different models that often converge back to Ising-like universality.
 We end by outlining the future directions in sociphysics research, highlighting the continued relevance of the Ising model in the
analysis of complex social systems.}




\maketitle

\section{Introduction}\label{sec1}
Sociophysics is a scientific discipline that employs ideas from statistical physics to analyze and interpret social phenomena. The roots of this approach can be traced back to the 17$^\text{th}$ century, when the English philosopher Thomas Hobbes suggested that societies are governed by laws similar to those in physical systems. This perspective was later echoed by several French philosophers, most notably by Auguste Comte in the 19$^\text{th}$ century, who expressed the view:

\begin{quote}
Social physics is that science which occupies itself with social phenomena, considered in the same light as astronomical, physical, chemical, and physiological phenomena, that is to say as being subject to natural and invariable laws, the discovery of which is the special object of its researches.
\end{quote}

The central argument was that, in sufficiently large populations, collective behavior emerges much like in thermodynamic systems, pointing to the existence of underlying universal laws. This idea gained further support as large-scale data on social behavior began revealing striking similarities signifying universal behavior. While individual humans possess free will, the observed emergent patterns suggest that such individual variations may act merely as small fluctuations within the broader collective dynamics.

\begin{figure*}[h!]
   \centering
    \includegraphics[width=\linewidth]{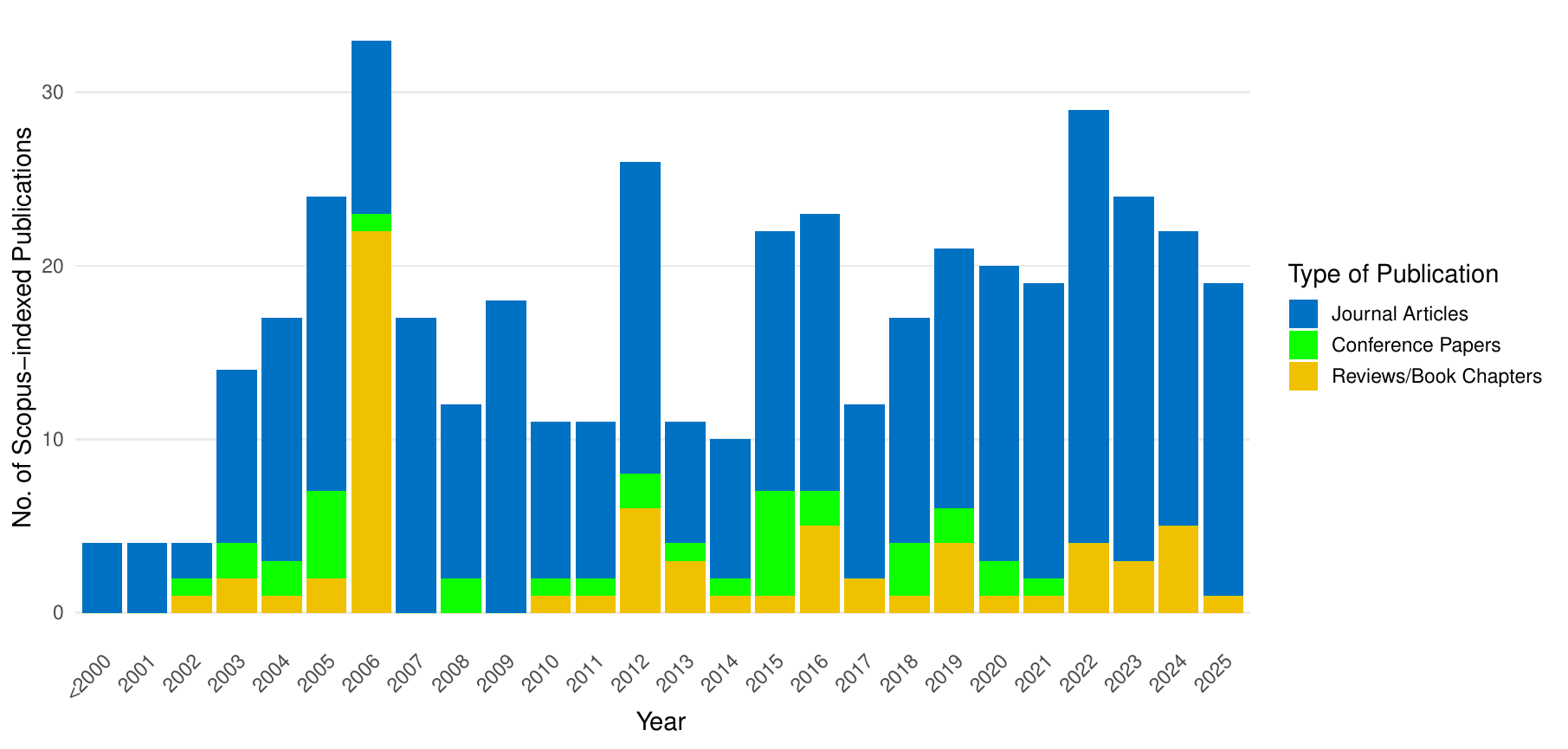}
    \caption{Bibliometric analysis of publications on \textit{Sociophysics} indexed by \textit{Scopus} from before 2000 to 2025 (June). The plot shows the number of publications per year, categorized into journal articles, reviews and book chapters, and conference papers. There is a noticeable increase in the number of publications, particularly in the last two decades, reflecting the growing interest in sociophysics.}
    \label{fig:biblio}
\end{figure*}

Sociophysics has gained considerable attention in the last few decades and its aim is to explain and understand  collective behavior in human societies. To highlight the relevance of sociophysics as an area of research we conducted a bibliometric analysis using \textit{Scopus} database with a keyword search of `\textit{Sociophysics}'. This query, performed in June 2025, resulted in a total of 443 publications so far, whose year-wise distribution has been shown in Figure \ref{fig:biblio}.
The bibliometric analysis shows a significant increase in sociophysics publications since 2005, with journal articles being the predominant format, indicating growing interest and consolidation in the field.

The Ising model \cite{newell1953theory,brush1967history}, originally developed to understand phase transitions in magnetic systems, has found a wide range of applications far beyond its original context. Its simplicity with the spins having binary states and  pairwise interactions provides a framework for modeling complex systems with binary state variables, including those found in social sciences. It is also possible to define the Ising  model  with interactions taking place over any range. The interactions can be extremely localised,  i.e., taking place with nereast neighbours only where a lattice picture works well  or  infinite ranged; the latter is applicable in cases of fully connected networks. The model is especially useful for studying social phenomena among various other fields. Sociophysics \cite{galam2008sociophysics,stauffer2013biased,sen2014sociophysics}, a field lying at the interface of physics and social behavior, often adopts models inspired by the Ising model to understand phenomena such as opinion dynamics \cite{pluchino2005changing,weisbuch2006social,sobkowicz2009modelling,crokidakis2012effects,galam2013modeling}, collective decision-making \cite{butt2014comparing,lorenz2018opinion,tsarev2019phase,tanimoto2019evolutionary}, social influence \cite{schweitzer2018sociophysics}, social segregation, game theory, several features related to 
finance and markets  \cite{sornette2014physics} and even in language adaptation \cite{stauffer2008social}. These phenomena often involve individuals or agents adopting binary states, representing contrasting opinions, behaviors, or decisions etc., thus making the Ising-like framework an ideal starting point for the analysis. In fact, one cannot but agree with Galam's statement \cite{Galam2012}

\begin{quote}
    Every person studying the Ising ferromagnetic model within the frame of modern statistical physics would envision an analogy with some social systems. It is a
very appealing universal model, which could apply to a large spectrum of social
situations.
\end{quote}

The reason why the Ising model has been so important in the context of social behavior is that, 
interacting individuals tend to increase their similarity in general.  It is normal to find  that opinions, cultural features, and languages are shared by large groups of people \cite{Sapiens}. Such similarity between interacting entities is analogous to ordered states in the Ising model, where, below the critical temperature, a majority of the spins share the same orientation.

The Ising model and its extensions have been used to represent social interactions and organization \cite{weidlich1971statistical,weidlich1991physics,weidlich2000sociodynamics,callen1974theory,montroll1974introduction,galam1982sociophysics,orlean1995bayesian} since a long time. Indeed, the analogy between magnetic polarization and opinion polarization was presented in the early 1970s by Weidlich \cite{weidlich1971statistical}, in the framework of “Sociodynamics”, and later by Galam et al. \cite{galam1982sociophysics} in a manifesto for “Sociophysics” in 1982. In that decade, several efforts towards a quantitative sociology developed \cite{schelling1971dynamic,granovetter1978threshold,granovetter1983threshold}, based
on models essentially indistinguishable from spin models.

In sociophysics, one of the key questions is how the  interactions and individual preferences lead to macroscopic societal behaviors. Many models have been proposed to capture this, with particular attention to opinion formation, where agents adopt different stances and influence each other over time. In this review, we explore various sociophysics models inspired by the Ising model, highlighting their ability to reproduce phenomena such as phase transitions, critical behavior, as well as non-equilibrium behavior. Although the equilibrium behavior may be Ising-like, often the dynamical behavior turns out to be different. We also examine models that incorporate dynamics more specific to social behavior, and discuss their relevance to real-world data, offering insights into how well these theoretical frameworks mirror observable social patterns.

In this review, we have focused on the application of the Ising model and its variants to several key areas within Sociophysics, including opinion formation, financial markets, segregation dynamics, language evolution, and the spread of COVID-19. Connection to some real world data, whenever possible, has been emphasised. Some data-driven Ising like models have also been proposed, which we discuss briefly in the last section.  
We have attempted to present an updated account of the theoretical
models, omitting the details for which one can refer to the original papers. In the process, we have limited the citations to works published in peer-reviewed journals and books.


\section{Brief history of the Ising model: equilibrium and nonequilibrium behavior}\label{sec2}

The Ising model,  formulated by William Lenz in 1920 and later solved in one dimension by Ernst Ising in 1925, was originally designed to describe phase transitions in magnetic systems. It models a lattice of discrete ``spins'' $\sigma_i$ (the quotes are put to emphasize that the concept of spins, which is purely quantum mechanical, was yet unknown at that time).   Each spin can take values $\sigma_i=\pm 1$, representing the magnetic moment at site $i$. The Hamiltonian for the system in its simplest form is given by: \begin{equation}
    H=-\sum_{ij}J_{ij}\sigma_i\sigma_j-h\sum_i\sigma_i.
\end{equation} Here, $J_{ij}$ is the coupling constant that represents the interaction strength between the $i$th and $j$th spin and $h$ is an external magnetic field. The interactions are taken to be symmetric, $J_{ij} = J_{ji}$ and  $J_{ij} =0$ if the $i$th and $j$th spins are not connected. In lattices, it is usually the nearest neighbors which are connected while on complex networks the connections may be long ranged. In the absence of an external field, the model encapsulates how local spin interactions lead to macroscopic phenomena such as spontaneous magnetization and phase transitions.

The one-dimensional Ising model, which Ising himself solved, showed no phase transition at finite temperature. However, the model gained prominence when Lars Onsager solved the two-dimensional case exactly in 1944, demonstrating that the system undergoes a phase transition at a non-zero critical temperature \cite{onsager1944crystal}. The significance of this work
 is reflected in a comment by Wolfgang
Pauli about the developments of theoretical physics during the second World
War: ``nothing much of interest has happened except for
Onsager’s exact solution of the Two-Dimensional Ising
Model" \cite{Bhatta}. This breakthrough paved the way for the Ising model to become a cornerstone of statistical mechanics, particularly in the study of critical phenomena and universality.

Since then, the model has been extended to various higher-dimensional systems \cite{aizenman1986critical}, and even generalized to more complex situations, such as quantum systems \cite{suzuki2012quantum}. Its versatility and simplicity made it an appealing framework not only for physical systems but also for understanding analogous behavior in biological, economic, and social systems \cite{sen2014sociophysics}. In sociophysics, the binary nature of the spins naturally maps to binary decisions, opinions, or behaviors, allowing for the exploration of collective dynamics in a wide range of contexts.

The Ising model, being a classical model, has no intrinsic dynamics. However, one can study the kinetics by taking the system  to  a configuration far removed from its equilibrium state.
Typically, the relaxation behavior and related features, often characterized by independent exponents, have become  topics of extensive research over the last few decades. 
The kinetic Ising model, where dynamical evolution takes place as the system is out of equilibrium, was introduced by Glauber in 1963 \cite{glauber1963time}. Such an evolution, where the system is expected to reach an equilibrium state after a very long time 
can be studied using different  
dynamical rules, necessarily obeying detailed balance. The standard rules are Glauber and Metropolis 
rules in which a single spin is tested to flip at any time step. 
In the Glauber scheme,  suppose the energy change due to the flipping of a spin is $\Delta E$.  Then the  spin  flip move is accepted  with a probability 
\begin{equation}
p = \frac{1}{1+e^{\beta \Delta E}}
\label{glauber}
\end{equation}
where $\beta = \frac{1}{k_BT}$, with $k_B$ the Boltzmann constant and $T$ the temperature.
The spin is flipped with probability 0.5 is $\Delta E $ is zero. Typically, in a Monte Carlo simulation, a spin is randomly picked up and $\Delta E$ and $p$ are calculated.  $p$ is compared to a random number $r$, if $p > r$, the spin is flipped. In the Metropolis algorithm \cite{hastings1970monte}, if the energy is lowered or remains same by the flip, the move is always accepted and otherwise the spin is flipped with a probability $e^{-\beta \Delta E}$. Another scheme is the heat bath dynamics where, after randomly selecting a spin (labeled by $i$), the  probability
\begin{equation} 
p_{hb} = \frac{1}{1+\exp (-2\beta h_i)}
\label{hb}
\end{equation}
is determined , where $h_i$ is the local field on the chosen spin. Then one sets  $s_i=1$ with 
probability $p_{hb}$. Note that the initial state of the spin does not affect $p_{hb}$ and 
the latter is simply defined in analogy with equilibrium systems.

\begin{figure}
\includegraphics[width=5.5cm,angle=270]{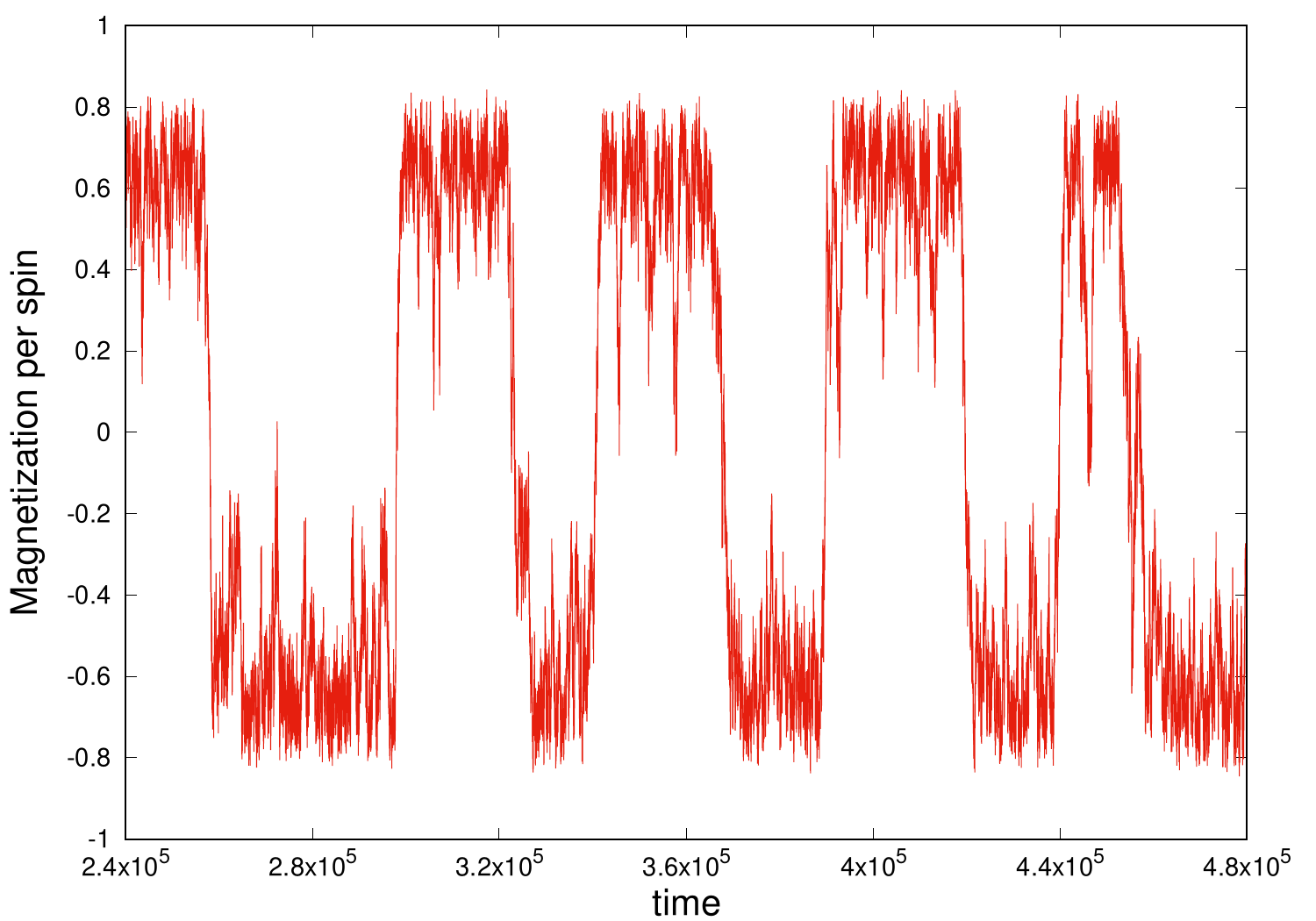}
\caption{Behaviour of the magnetisation in the two dimensional Ising model near criticality as a function of time as obtained using numerical simulations. Typically, it oscillates between positive and negative value (Figure courtesy Amit Pradhan)}
\label{ising-oscill}
\end{figure}

As an example of the study of the Ising kinetics, it is shown in Fig. \ref{ising-oscill}, how the magnetization, given by the average sum of the spins, oscillates between positive and negative values close to the critical temperature in the two dimensional Ising model. 

The time evolution of the Ising model, where a single spin is updated at every time  step, can be used in the social context when the states of the individuals are given by  binary values and it is needed to check whether they will change their states in a single time step.  The probability of this depends on the ``temperature'', used as a parameter, which can be interpreted  as  noise.

Another scheme, namely the Kawasaki kinetics, works for systems with conservation, i.e., where the number of up and down spins are constant individually. Here two oppositely oriented spins interchange their
positions on the route to equilibrium. As we shall see, this becomes useful for situations
where social segregation takes place. 

The zero temperature Ising-Glauber model often becomes most relevant with respect to social dynamics. 
Here, the updates are quite simple: if a single spin flip increases the energy, nothing is done. In case the energy decreases, the move is accepted. When the energy remains constant, the spin is flipped with probability 1/2.  Of course, the concept of energy is absent in social models, still, equilibrium states and deviations from that can often be identified. 

\begin{figure}
    \centering
    \includegraphics[width=0.8\linewidth]{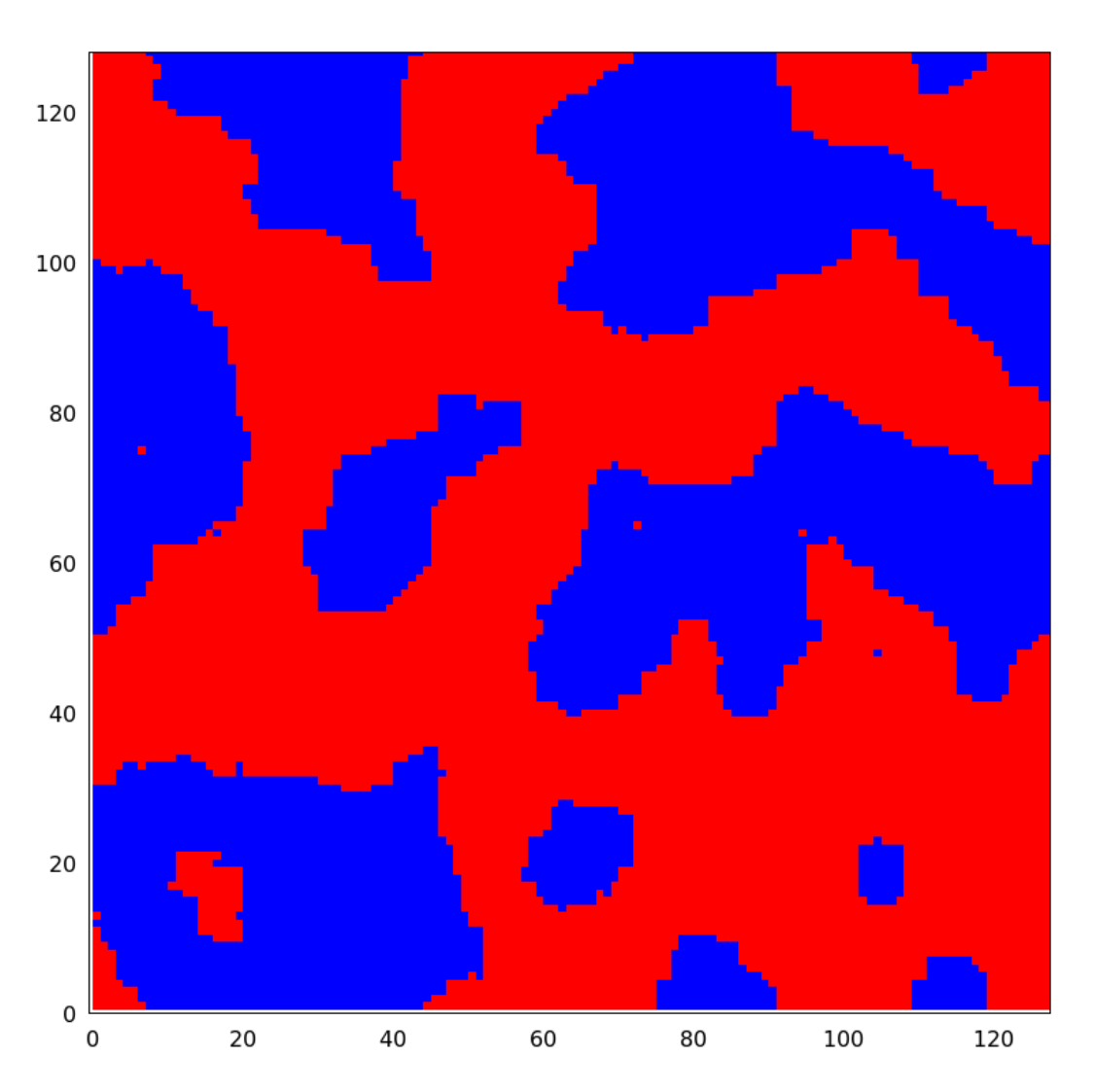}
    \caption{A snapshot of the two dimensional Ising model undergoing a quench to temperature $T < T_c$.(Figure courtesy Amit Pradhan)}
    \label{ising-dynamics}
\end{figure}

 When the interactions are more complicated as in a spin glass, where the interactions $J_{ij}$ are quenched random variables having both  positive and negative values, the system may not reach the equilibrium state at all using the standard algorithms. The system, in such cases, may reach a nonequilibrium steady state or an active state where the dynamics can continue forever. This can also be true in case of social phenomena for which equilibrium is not defined and 
 one can only study the time evolution of the system starting from arbitrary initial conditions. Both the equilibrium and nonequilibrium behavior are thus meaningful for social scenarios which makes Ising and Ising-like models so important to study even after a century of its inception \cite{ising2017fate,macy2024ising}.

\section{Sociophysics models: Connection with Critical Phenomena}\label{sec3}

At its core, sociophysics leverages models inspired by physical systems to describe and predict emergent patterns in opinion dynamics \cite{Rainer2002-RAIODA,pluchino2005changing,castellano2009statistical,acemoglu2011opinion,xia2011opinion,galam2013modeling,das2014modeling,sirbu2017opinion,schweitzer2018sociophysics,lorenz2018opinion}, voting behavior \cite{gwizdalla2008gallagher,President,biswas2021block,galam2021will}, crowd movement \cite{nagel1992cellular,helbing1995social,schreckenberg1995discrete,fukui1999jamming,burstedde2001simulation,perez2002streaming}, and more. A key advantage of these models is their ability to replicate complex social phenomena using relatively simple microscopic rules, involving few parameters, that govern individual interactions. 


One of the most intriguing aspects of sociophysics models is that in large number of cases they  manifest  
phase transitions driven by the parameters of the models. Such behavior, especially when the phase transition is continuous in nature, can be studied with the available tools of statistical physics. Critical phenomena, extensively studied in physical systems \cite{stanley1971phase,hohenberg1977theory,herbut2007modern,ma2018modern}, refer to the dramatic changes in system-wide behavior that occur when a parameter, such as temperature in a magnetic system, approaches a critical point. In the context of sociophysics, such phenomena arise when a small change in individual behavior can trigger large-scale shifts in societal states — whether it be the rapid spread of an opinion, large fluctuations in stock prices, a sudden shift in electoral outcomes, or the tipping point in a social movement. Opinion dynamics models such as the majority rule model \cite{coughlin1990majority,eraslan2002majority,krapivsky2003dynamics,chen2005majority,crokidakis2015inflexibility}, the Sznajd model \cite{sznajd2000opinion,stauffer2002sociophysics,slanina2003analytical,sznajd2005ActaPhysica,sznajd2011phase,sznajd2021}, and several variants of the voter model \cite{castellano2009nonlinear,nyczka2012phase,timpanaro2014exit,chmiel2015phase,mobilia2015nonlinear,vieira2018threshold} have demonstrated the occurrence of phase transitions in opinion dynamics, showing that societies can transition between consensus and polarization states depending on factors like interaction rules or external influences. In financial systems, such criticality occur in the form of bubbles or crashes.

The notion of spontaneous symmetry breaking, central to critical phenomena in physics, plays a key role in sociophysics. Just as magnetic systems spontaneously choose a magnetization direction below the critical temperature, social systems can spontaneously break symmetry by, for example, converging on a dominant opinion or behavior, even in the absence of external bias. This spontaneous alignment in social systems often resembles a continuous phase transition, where the system gradually transitions from a disordered (e.g., disagreement in opinion formation models) state to an ordered (e.g., consensus in opinion formation models) state as the system's control parameter(s), such as the strength of social influence, crosses a critical threshold. 


Critical exponents, which describe how system properties behave near the critical point, have also been investigated in sociophysics models. The values of the exponents help to identify the universality classes, which, in physical systems, is dependent only on the spatial dimension and the symmetry of the so called order parameter. These concepts allow researchers to group different sociophysics models into broader categories based on their collective behavior rather than their specific rules. For instance, it has been shown that certain sociophysics models exhibit the same critical exponents as classical spin systems, indicating that despite their differences, the underlying mathematical structures governing their behavior near the critical point are similar. 

While much of the focus in sociophysics has been on modeling phase transitions, researchers have also explored other interesting features, such as persistence, exit probability and synchronization, the latter in the special case when the society has a modular structure.

Persistence $P(t)$ refers to the probability that an individual agent, starting in a certain state, will remain in that state up to the time $t$ of observation.
Exit probability $\mathcal{E}(x)$, another important quantity, is the probability that a system will end up in one of the two possible absorbing states, such as complete agreement or disagreement in a population, where $x$ is the initial fraction of agreeing agents. This metric is particularly useful in predicting long-term outcomes in opinion dynamics and voting models, where the final state of the system is often one of total polarization or consensus. The time to reach consensus $\tau$ is also important, as a system can be locked in a metastable state for considerable duration such that the properties manifested will correspond to that observed at finite times. $\tau$ could be measured as the average number of time steps required by the system to reach a consensus.


\section{Models of Binary Opinion Dynamics}

As already mentioned, the maximum application of Ising and Ising-like models has been in the field of opinion or decision forming.  
In many cases, like in a public debate/referendum  or meeting, one has a ``yes/no" situation such that 
the opinion states can be regarded as  binary. Initially the states may be random but typically,
after several rounds of arguments and discussions, an agreement of the majority of the agents can be reached. This is simply an example how emergent opinions can be formed in an iterative way, a scheme followed in almost all opinion dynamics models. Of course there may be situations where more than two discrete opinions can exist or continuous  values of opinions are possible but for the present article, we restrict to systems with binary states that can
be represented by Ising spins.

\begin{figure}[h!]
    \centering
    \includegraphics[width=\linewidth]{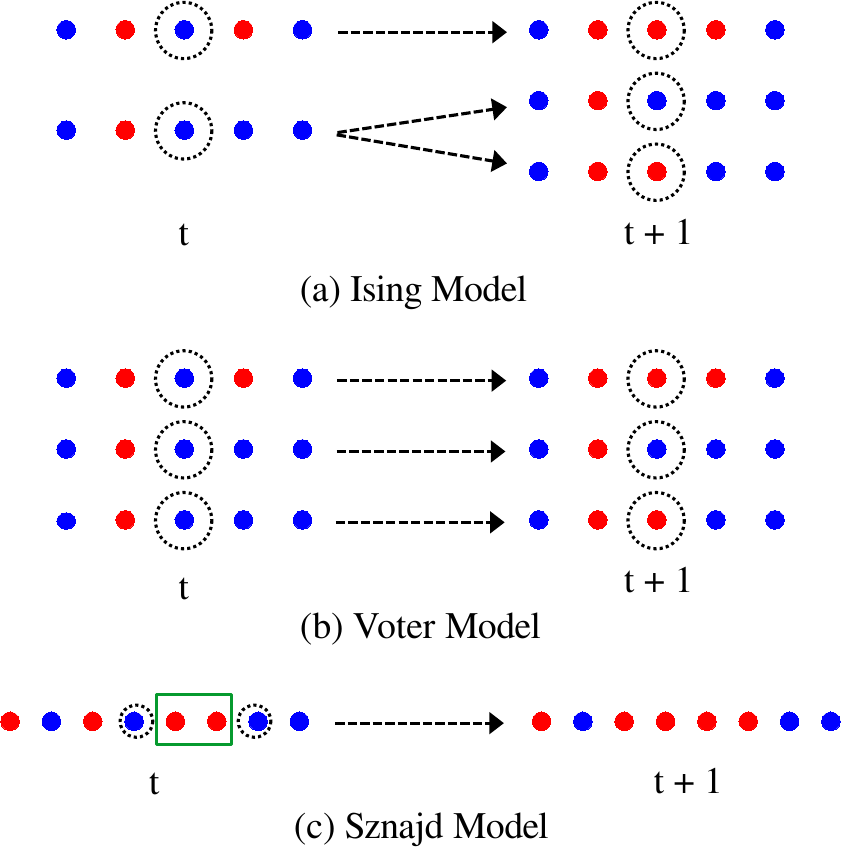}
\caption{Schematic representations of the evolution of one dimensional (a) Ising model (b) Voter model and (c) Sznajd model, from time $t$ to $t+1$. The red and blue circles denote the binary states, such as `up' and `down' spins, or `yes' and `no' opinions etc. The focal agents are denoted by dotted circles. (a) For Ising model under Glauber dynamics, when the focal agent is surrounded by opposite type of agents (\textit{top panel}), it gets flipped. But when it is surrounded by a similar and an opposite type of agent (\textit{bottom panel}), it flips with probability $1/2$. (b) For the voter model, the focal agent randomly chooses one of its neighbours, and adopts its state. The dynamics are effectively the same as that in (a). (c) For Sznajd model, a pair of agreeing neighbours, denoted by green rectangular box, influences its neighbours to adopt the same opinion.}
\label{fig:ising_1D}
\end{figure}

One of the simplest binary opinion dynamics models is the Voter Model \cite{holley1975ergodic}, where each agent adopts the state of a randomly chosen neighbor at each time step, irrespective
of her own state at that moment. This model captures the mechanism of social influence and consensus formation through local imitation and 
models the phenomena of things going `viral'. Despite its simplicity, the voter model exhibits rich dynamics, including coarsening phenomena and the possibility of reaching absorbing consensus states.  In one dimension, the voter model and the kinetic Ising model
using Glauber dynamics (discussed in sec \ref{sec2}) are equivalent as demonstrated in Figs. \ref{fig:ising_1D}(a) and \ref{fig:ising_1D}(b). In higher dimensions, however, the mechanism of coarsening is curvature driven in the Ising model while it is interfacial noise driven in the Voter model. 

Consensus in the voter model is  reached only for $d \leq 2$ for an infinite system.
However, 
in finite lattices consensus is reached asymptotically; here the system may get locked in a metastable state for a transient time,
but eventually can escape from it.
  A random fluctuation can drive the
system from the metastable state to one of the two ordered absorbing states.
The time $\tau$ required to reach consensus
in finite systems varies with the system size $N$, in one dimension, where the dynamic exponent for coarsening is 2, it varies as
$\tau \sim N^2$ as in a Ising system,
in two dimensions $\tau \sim N \log(N)$ while for $d=3$ it varies linearly with $N$.

Terming the two opinions as $A$ and $B$ (quantified as 1 and -1), since an $AB$ pair goes to $AA$ or $BB$ states with equal probability, it is obvious that the order parameter (total opinion) will be conserved when an ensemble average is taken. Taking $x$ as the initial proportion of opinions  with value 1, the  exit probability $\mathcal{E}(x)$ must therefore satisfy
\begin{equation}
\mathcal{E}(x) - [1-\mathcal{E}(x)] = 2x -1,
\label{exit-voter}
\end{equation}
 implying $\mathcal{E}(x) = x.$ This is true for the voter model in any dimension.

Another important model is the Majority Rule Model \cite{coughlin1990majority,eraslan2002majority}, where groups of agents adopt the majority state within their local neighborhood. This model emphasizes the role of group interactions and can lead to faster consensus compared to the voter model. It has been used to study phenomena such as opinion polarization and the impact of group size on consensus dynamics \cite{krapivsky2021divergence}.



In two dimensions, the Ising, Voter and the majority rule model can be achieved using a generalised parametric model with two parameters \cite{de1993nonequilibrium}. The phase space is spanned by the two parameters, $z$ and $y$.
In such a generalised model, the probability $p_i$ that the spin $\sigma_i$ will flip is given by \begin{equation}
    p_i(\sigma)=\frac{1}{2}\Big[1-\sigma_iF_i(\sigma)\Big].
\end{equation} Here, $\sigma=\sum_{\delta}\sigma_{i+\delta}$ is the sum of the four neighbouring spins of $\sigma_i$, which can take values $\pm 1$. The function $F_i(\sigma)$ is defined as \begin{eqnarray}
    F(0)=0,F(2)=F(-2)=z\label{eq:z}\\
    F(4)=F(-4)=y\label{eq:y}
\end{eqnarray} where the parameters are restricted to $z\leq 1$ and $y\leq 1$. The points $z=1,y=1$ and $z=0.5,y=1$ correspond to the Ising model at zero temperature and the voter model, respectively. On the other hand, the line $z=y$ corresponds to the majority rule model.


A  model based on the 
the concept of social validation \cite{sznajd2000opinion}, namely, the Sznajd model also bears some interesting similarities and dissimilarities with the Ising model. Here a pair of agreeing neighbors can influence their surrounding neighbors to adopt the same opinion, as shown schematically in Fig. \ref{fig:ising_1D}(c) in one dimension. This mechanism reflects the idea that unified opinions within a small group can have a persuasive effect on others, leading to the spread of opinions in a society.
The Sznajd model has the same two absorbing states as in the Ising model.
The two models also  have identical
exponents associated with domain growth and persistence behavior during coarsening \cite{behera,stauffer1}.
The two models can be differentiated on the basis of flow of information; in the Ising model,
the information flows inwards to a particular agent while in the Sznajd model, it is the opposite.
 A few  quantities related to the dynamics were shown to be different for generalised  models with inflow and outflow dynamics,
where a suitable parameter associated with the spin flip probability was introduced \cite{godreche,krupa}.
The Ising Glauber and Sznajd models can be obtained by choosing specific values
of the parameters in the generalised models with inflow and outflow dynamics respectively. The exit probability, in particular, depends on the direction of flow of information \cite{oarna_ps2014}.

\begin{figure}
    \centering
    \includegraphics[width=0.8\linewidth]{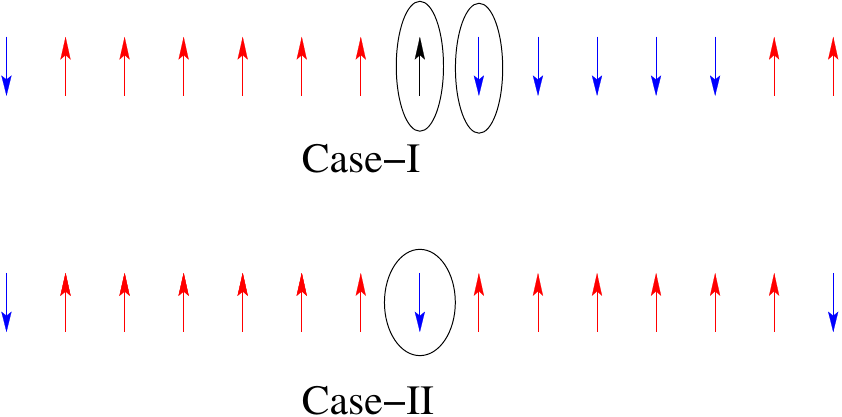}
\caption{The updates in the model with domain size dependent dynamics.  In case I, the focal spins (circled) are located at the boundary of two domains of opposite spins; the larger domain will determine the state. In this figure, the  spin shown in black will not flip while the blue will flip. In case II, the spins sandwiched between two spins of opposite orientation will always flip.}
\label{BS}
\end{figure}

These binary models often exhibit phase transitions between ordered (consensus) and disordered (diverse opinions) states, depending on parameters such as the size of local groups or the influence of external fields \cite{sen2014sociophysics}. The critical behavior observed in these models draws direct parallels to the Ising model, reinforcing the applicability of statistical physics concepts in sociophysics.

A one dimensional model, where the sizes of the neighboring up and down domain are taken to represent social pressure, has also been considered \cite{sohma_PS2009} where the dynamical rules differ from the zero temperature Glauber scheme as demonstrated in Fig. \ref{BS}. In this case, only an agent sitting at the boundary of two domains of positive and negative opinions can change her opinion. The size of the two domains on either side are compared, and the agent adopts the opinion of the larger domains. The absorbing states are still the all up/down states, however, the dynamical evolution is much faster such that the exponents associated with domain growth and persistence are markedly different from those of 
the Ising model. 


 The Ising Hamiltonian with both negative and positive interactions was considered in \cite{afra2013} where party affiliations directly governed the interactions. Such signed interaction was also considered on complex networks in a binary opinion dynamics model where an Ising Hamiltonian was used \cite{LI-ising-neg} and the Metropolis algorithm was used to update the states. A phase transition was found above a critical value of the fraction of negative interactions in both cases. The Ising-PageRank model \cite{FRAHM2019121069}, a recent extension of the Ising framework, combines local binary interactions with the global structure of directed networks. Each node carries a spin-like binary state, and their interactions are governed by a doubled Google matrix that mimics spin-spin coupling. The model shows how a small elite subset of high PageRank nodes can exert a dominant influence on overall opinion formation, effectively acting as influential agents that drive large-scale polarization in real-world networks such as Wikipedia and the Oxford University web graph. More recently, the Ising model has been applied to clustered networks to investigate how community structure influences opinion dynamics, showing that metastable states and critical configurations play a central role in collective opinion shifts \cite{baldassarri2023ising}.

\section{Ising  model applications to Econophysics}


Just like the opinion dynamics models where binary opinion states are expressed in terms of Ising spins $\sigma= \pm 1$, in the financial scenario, one can associate similar variables. There has been a number of research works relevant to stock markets and phenomena like bubbles and crashes  can be described using the concepts of the Ising model. 

A detailed account of the development of microeconomical models using statistical mechanics is available in \cite{sornette2014physics}. Essentially one  
considers $N$ traders in a social network, whose links represent the communication channels between the traders.  The traders can either buy or sell one
asset at a time dependent price. In the simplest version of the model, with time considered as discrete, each agent can
either buy or sell only one unit of the asset at any time step. This is quantified by the buy
state $s_i = 1$ or the sell state $s_i = -1$. Each agent can trade at time $t$ and the decision to buy or sell is 
 based on all previous information up to $t -1$. It is assumed that the agents' decisions are based on rational optimization of the expectation of the payoffs of the agents. The earliest attempts made by Brock and Durlauf (1999,2001) \cite{BD1999,BD2001} involved the construction of  a stylized model of community
theory choice based on the agents' utilities. This model  contains a term quantifying the
degree of homophily which, in a context of random utilities, leads to a 
formalism essentially identical to the mean field theory of magnetism.
In \cite{Roehner-Sornette2000}, another  model was constructed 
in which an equation for the price variations was proposed and it was shown that an agent can maximise her profit provided $s_i(t)$ follows an equation which becomes identical to that followed by the spins under  Glauber dynamics under certain conditions.

In a number of works, the local field which determines the state of a  trader was 
formulated which contained  terms involving Ising-like interactions.
In several of these works the aim was to explain bubbles and crashes which are extreme events in finance characterised by the sudden rise or fall of stock prices. The existence of an underlying Ising phase transition was found to be 
responsible for such phenomena.
In  \cite{bornholdt2001expectation},
the local field   $H_i(t)$ was defined in terms of an interaction term and a global coupling to the magnetisation
as follows
\begin{equation}
H_i(t) = \sum_{j=1}^N J_{ij} s_j(t) - \alpha C_i(t) \frac{1}{N}\sum_{j=1}^N s_j(t).
\end{equation}
The first term is nothing but a Ising interaction term with $J_{ij} = J$ if $i$ and $j$ are connected and zero otherwise and represents the interaction strength. The second term is related to the global magnetization with $C_i$ representing the strategy of the $i$th agent coupled to the magnetization and 
 $\alpha > 0$ is a constant. The sign of $C$, if positive, implies a tendency to follow
the minority as is done in many social circumstances while $J > 0$ favors ferromagnetic interaction.
The probability $p$ that $s_i =1$ was given in terms of $\beta$, a responsiveness parameter, which governs the update rules for the states. The local evolution rule, in a system of $N$ agents was described as:
$ s_i(t+1) = +1$ with probability  $p = 1/[1+\exp(-2\beta H_i(t)],$
and with probability $(1-p)$, $s_i(t+1) = -1.$ This is  nothing but the heat bath dynamics where the responsiveness parameter corresponds to the inverse temperature used in Eq. \ref{hb}. 

An interacting agent based model of speculative activity was presented using an infinite dimensional Ising model in which a simpler expression for the field was used with an interaction term and an external field term, both time dependent, in \cite{KAIZOJI2000493}.
Here traders get influenced by the investment attitude of other traders, which
gives rise to regimes of bubbles and crashes interpreted as due to the 
collective behavior of the agents at the Ising phase transition and in the ordered
phase.

Numerous other models have been proposed subsequently with an Ising like  effective field  acting on the agents which incorporate additional features like memory, 
stylized facts, noise, actual price dependence, learning, effect of mass media  etc. Bubbles and crashes have also been investigated using Ising like models on different topologies.  We refer the reader to \cite{sornette2014physics,zha2020} for more detailed descriptions and references.

Apart from the above mentioned features, another issue of interest 
which can be compared to an Ising model feature is stock market prices. The latter is known to have a fat tailed distribution.
In \cite{chowdhury1999generalized} an Ising-like stock market model for financial markets using interacting ``super-spins" to capture trader behavior was formulated, demonstrating the emergence of market bubbles, crashes, and fat-tailed distributions of stock price variations through empirical analysis. 

Volatility is a measure of how uncertain the future price of a given stock is. 
A microscopic model of financial markets consisting of many interactive agents was considered in \cite{Kra-Holyst}  with global coupling and discrete-time heat bath dynamics, similar to random Ising systems where the interactions between agents change randomly in time. In the infinite population limit, the  time series of price returns showed chaotic bursts resulting from the emergence of attractor bubbling or on-off intermittency, resembling the empirical financial time series with volatility clustering. For suitable choices of the model parameters, the probability distributions of returns exhibit power-law tails with scaling exponents close to the empirical ones.
In \cite{harassetal2012}, the time evolution of an interacting system where an external driving force and noise are present was considered. The dynamical equation is equivalent to the Ising-Glauber model provided the
random variables follow a particular distribution. The volatility is then estimated in terms of the fluctuations in the time averaged magnetization. It was found that there exists noise induced volatility indicated by strong fluctuations in the collective dynamics of bistable
units. An analytical theory of a stochastic dynamical version of the Ising model on regular and random
networks was also conducted which demonstrated the ubiquity and robustness of this phenomenon. Using  the spin model of \cite{bornholdt2001expectation}, the cumulative distribution of log-returns was shown to exhibit scaling with exponents steeper than 2 in \cite{KAIZOJI2002441}.


\section{Three state models of opinion and financial dynamics: Ising connections}

Although the binary models of opinion dynamics are closest to the Ising model,
opinions can have multiple values as well as continuous ones. However, even in such cases, 
there is usually a finite number of states to which the system evolves, e.g., either a 
consensus state with all opinions same or polarisation occurs when the opinions get grouped 
into two values. 
Kinetic exchange models of opinion formation form a special class of opinion dynamics model where there is a pairwise interaction of two agents at every time step.  
We discuss a special case of a kinetic exchange model of opinion dynamics popularly called the BChS model \cite{BChS}, where one can consider 
three opinion states   $s_i = \pm 1, 0$ of the $i$th agent. The zero state   may correspond to   the case where the agent is neutral.  Here, the  $i$th agent  changes her opinion after interaction with the 
$j$th agent following the equation
$$s_i(t+1) = s_i(t) + \mu_{ij} s_j(t),$$
where the $\mu_{ij} = \pm 1$ represent interaction strengths, chosen randomly in an annealed manner. The opinions are  bound, $s_i \in [-1,1]$ so that the values are adjusted to 1 (-1) if it exceeds 1 (becomes less than -1). The interactions can be both positive and negative; a negative value is taken with probability $p$. It is seen that $p$ acts as a noise which drives a phase transition between an ordered state and a disordered state. 

Interestingly, on two and three dimensional lattices, as well as in the mean field limit,
the critical exponents are identical to those of the Ising model \cite{sudip-arnab2016}. However, the dynamical
behavior is different. Apparently there are two time scales and some long lived metastable states
due to the presence of the zero states. This model reproduces the pattern of opinion changes
observed in the different surveys conducted before (and after) Brexit \cite{BMS}.

Another application of the BChS model has been for US Presidential election, to estimate the probability that the popular winner loses after the electoral college procedure, which is like a coarse graining. The results from both the Ising model and the BChS model both predict that 
such a scenario can occur with a finite probability, but the latter model gives a better agreement with the real data \cite{President}.


To model a different aspect in economics, namely the expectation of business managers about their business prospects, a three state model was proposed 
 \cite{hohnisch2005socioeconomic}.  This work was based on an actual survey, where the business managers' expectations were put in 
 three categories: negative, neutral or positive. As in the BChS model, in the model,  to each
agent,  there is associated a variable $s$ which can take the values 
$s = \{-1, 0, 1\}$ corresponding to the three categories.
A quantity analogous to the interaction energy is defined which is minimum when two neighbors are aligned. Using the Glauber dynamics, the simulation showed good agreement with the 
 empirical time series of the fractions of particular expectation types (negative, positive and neutral).
It was noted that in this three state model, 
there is a tendency for either a strongly pessimistic or strongly optimistic opinion to form,
similar to the Ising simulations shown in Fig. \ref{ising-oscill}. 



\section{Social Segregation and Ising model}


Social segregation, with respect to different features like race, religion or ethnic origin, even without external forces (e.g. political), has been observed in different places of the world. In order to explain such segregation, 
Schelling \cite{schelling1971dynamic} proposed a model which can be regarded as a paradigmatic 
model of such a phenomenon. In the simplest case, an individual can
belong  to any one of the two groups designated by  A and B,  assuming
that they occupy the sites of a square lattice. If $n_A $
and $n_B$ denote the number of neighbors of a particular agent belonging to 
group A, say, then in the simplest model, the person will be unhappy if $n_A < n_B$. An unhappy agent will move to an empty position in an attempt to increase the happiness factor.

\begin{figure}
    \centering
    \includegraphics[width=\linewidth]{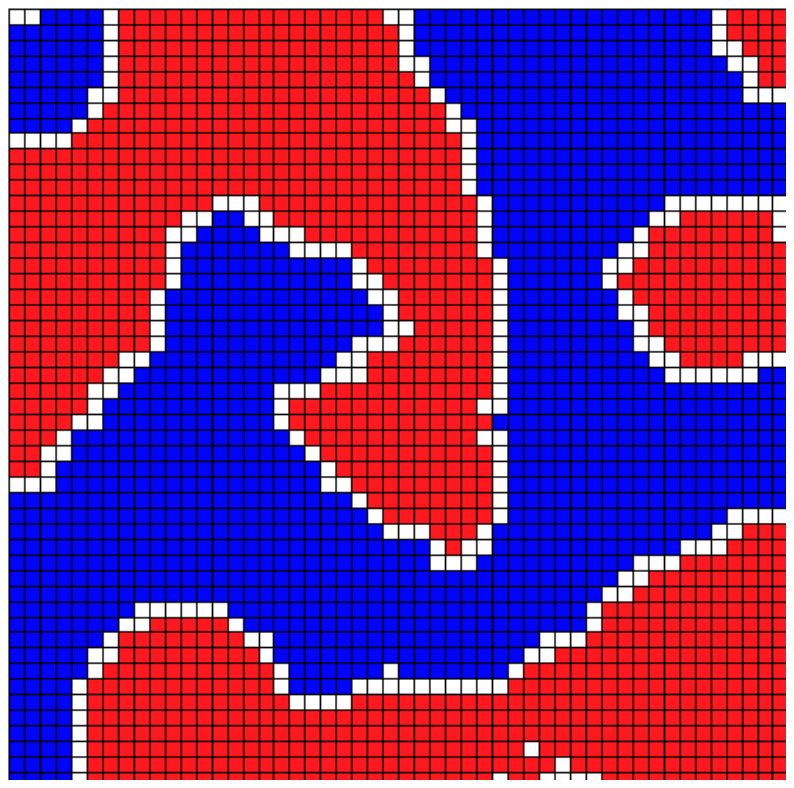}
    \vskip -5cm
    \caption{An example of the steady state using Schelling's model where tolerance limit is 70\%. Red and white colors represent two agents belonging to two groups, whites are empty spaces (figure generated using  http://nifty.stanford.edu/2014/mccown-schelling-model-segregation/)}
    \label{schelling}
\end{figure}

It was shown by physicists \cite{stauffer2007ising,muller2008inhomogeneous}, that such a segregation can be easily obtained in the Ising model, at a finite temperature below the critical point, when domains of up spins and down spins will grow from an initially disordered state. This resembles a segregated society, with either the up or down spins forming a majority keeping the absolute value of the magnetisation constant. In the social scenario, 
the person belonging to group A (B) can be assumed to have spin state $s=1$ $(-1)$. 
In the Ising model, the local field at a site determines the probability of the 
site having spin $\pm 1$. In context of the social segregation, the probability 
$p(A)_i$ ($p(B)_i)$) of the $i$th site belonging to group A (B) is dependent on  a similar field given by  $h_i=-(n_A-n_B)_i$ such that 
\begin{eqnarray}
p(A)_i & =& \frac{e^{\beta h_i}}{e^{\beta h_i} + e^{-\beta h_i}}\\
p(B)_i & =& \frac{e^{-\beta h_i}}{e^{\beta h_i} + e^{-\beta h_i}},
\end{eqnarray}
which is nothing but the heat bath algorithm of Ising model (eq. \ref{hb}). 
Here the scaled temperature $=1/\beta$ acts as a social temperature that can be interpreted as tolerance in the context of segregation. The original Schelling model corresponds to zero tolerance, i.e.,  $T=0$. A typical evolution of the system for particular parameter values are shown in Fig. \ref{schelling}, note the similarity with the Ising dynamics at a temperature less than $T_c$ shown in Fig. \ref{ising-dynamics}.

However, in reality, the number of up and down spins should remain constant if the feature on which the segregation is based cannot be changed (for example, religion can be changed but race cannot be). In that case, the 
Kawasaki dynamics, in which the unlike neighboring spins can change position such that the total spin remains same, can be used. The  probability that such an exchange of position occurs depends on  the change in the  happiness factor of both the agents and a temperature factor. In \cite{ghetto-kawasaki}, the initial concentration and the temperature were 
varied to show that the Kawasaki scheme indeed leads to ghetto formation for certain ranges of the parameters used.

The Schelling model has more connection with physics as it has resemblance with hydrodynamics models as well, but that discussion is outside the scope of the present review.

\section{Similarities between Language and  Ising dynamics}


In certain social phenomena, Ising model has not been directly applied but a striking similarity has been found in certain behavior of the social and Ising dynamics. One of these relates to language dynamics. In 
 \cite{nettle1999rate}, a study was made  using numerical simulations to find out how one language (or language feature), which is spoken by everyone,
can be replaced by another language,  without any external force or bias.  Here the agents, located on a lattice, are required to learn linguistic terms with two variants of each. The adaptation of a particular variant is determined by the total impact on the learner,
which depends on a number of factors like  the number of individuals using the variant, the geographical distance separating the individuals and their status averaged.  
It was found that the fraction of population adapting one variant showed irregular oscillations  which can be compared to the oscillations in the Ising model  between positive and negative spontaneous
magnetizations in the Ising model (Fig. \ref{ising-oscill}). It was also observed  that the rate at which the majority switches to the other  language decays as the 
population size  increases. Such a feature is known to be present in the Ising model;  the rate of switching between positive and negative magnetization 
decays exponentially with increasing linear dimension of the lattice \cite{stauffer2008social}.
However, the model considered in \cite{nettle1999rate} contained far more intricacies like presence of contrarians, socio-economic status of the agents etc. The topic of language dynamics has been studied extensively, here we mention only the similarities with the Ising dynamics.


Another interesting feature of any language is the ordering of words in a sentence. The structure of a sentence is characterised by the way the different words like verbs and adverbs etc are positioned in the sentence. This ordering varies from one language to the other, for example, Japanese, Korean and Tamil follow one particular order while in Thai, it is completely 
different. 
Hence, it appears that there can be two possible stable states and the world fluctuates between these two structures like the Ising model on a  finite lattice as shown in Fig. \ref{schelling} 
\cite{language2}.

\section {Game-theoretic models of Ising dynamics}


In standard game-theoretic frameworks, agents select strategies independently, without access to others’ decisions, while their payoffs depend on both their own and others’ choices. When only two strategies are available, these can be naturally represented as Ising spin states. For two-player interactions, the payoff structure is typically captured in a $2 \times 2$ matrix. A well-known example is the Prisoner’s Dilemma, where each player must choose between cooperation and defection.



Such games are often played iteratively, with each agent seeking to optimize their own payoff. As the interactions evolve, the system may converge to an equilibrium state, defined in different ways. A prominent example is the \textit{Nash equilibrium}, where no individual can gain by changing their strategy unilaterally.


Galam et al \cite{galam_game} studied generic $2 \times 2$ games and demonstrated that the Ising model provides a useful framework for analyzing such scenarios. In the Ising model, equilibrium corresponds to the minimum energy configuration, which can be reached through standard dynamical evolution, often referred to as Hamiltonian dynamics. By associating the elements of the payoff matrix with the energies of the four possible spin configurations of a two-spin Ising system, it was shown that the Ising model can reproduce all game-permitted outcomes under the condition of Nash equilibrium.


Ising-type Hamiltonian dynamics has also been employed to investigate equilibrium behavior in games involving more than two players. In the three-player public goods game, the equilibrium fraction of cooperating agents was shown to correspond to the expectation value of a thermal observable analogous to magnetization \cite{adami2018}. Furthermore, a transition between cooperative and defective behavior was identified, mirroring a phase transition in one-dimensional Ising systems with long-range interactions.


In the context of the Battle of the Sexes game, in \cite{Correia_game} it was demonstrated that equilibrium leads to correlated choices among players, with these correlations exhibiting an Ising-like structure. The analysis involved two or three agents on small networks. The correlation probabilities could be mapped onto an extended Ising model incorporating site-dependent fields and, in the three-agent case, a three-body interaction term. Similarly, Ising-based games on graphs were explored  in \cite{Leonidov-game}, introducing a noise component to account for the stochastic effects.
Potential games, where all players' strategic incentives are derived from a single global potential function, were investigated by modeling the system using the Ising Hamiltonian for large populations in \cite{tejasvi-game}.

\section{Modeling COVID-19 spread with Ising dynamics}

The spread of infectious diseases inherently involves patterns of social interaction, making it a natural candidate for modeling within the Ising framework. During the COVID-19 pandemic, several models were developed that assign $\pm 1$ states to represent infected and non-infected individuals, with interactions defined according to variants of the Ising Hamiltonian on different lattice structures.

In \cite{covid-mello}, multiple models to study the spread of COVID-19 were explored, including one based on the Ising model and other concepts from statistical physics. In this approach, agents occupy nodes of a Bethe lattice, with interactions depending on their current states. The model was able to retrospectively reproduce infection trends not only for COVID-19 in China but also for earlier pandemics such as Ebola, SARS and influenza A/H1N1.

A related model implemented on a square lattice showed qualitative agreement with the temporal evolution of infections \cite{Covid19cris}. The influence of lockdown measures was captured in \cite{covidsumour} by associating the Ising model's temperature parameter with the level of social interaction, where lower temperatures indicated more restrictive conditions. In another approach, a deep learning model based on Ising dynamics was developed for spatiotemporal prediction of COVID-19 hospitalizations  \cite{Covid19gao}, where model parameters were used from the real-world data.

\section{Conclusion and Outlook}\label{sec-conclu}

In summary, this review highlights a unifying theme across diverse sociophysical domains, from opinion dynamics and financial market fluctuations to social segregation, language dynamics, and epidemic spreading (e.g. COVID-19). In each case, simple binary-state agents interacting locally can spontaneously produce emergent collective behaviors strikingly analogous to phase transitions in physical systems. This observation marks the utility of the Ising model framework as a minimal yet powerful tool for understanding complex social phenomena. Despite its simplicity, the Ising paradigm captures how micro-level interactions (peer influence, imitation, etc.) can trigger macro-level outcomes such as consensus formation, market crashes, segregation patterns or contagion outbreaks, often with identifiable critical points or `tipping' phenomena. By drawing direct parallels to statistical physics, Ising-inspired models provide a coherent formalism to analyze sudden societal shifts and order–disorder transitions, illustrating the value of interdisciplinary approaches in social dynamics.

A key test of any modeling approach is its ability to accurately describe and predict empirical phenomena. Sociophysics models, inspired by statistical physics frameworks such as the Ising model, have generated considerable interest due to their conceptual simplicity and versatility. However, for predictive purpose, one has to be careful as the the parameters affecting collective decisions can be many in the real world scenario. A recent work \cite{vendeville2025voter} shows correspondence between online opinion data and a Voter model with zealots. In \cite{President}, the cases when the popular US presidential candidate lost after counting by electoral college was done, had been considered and the Ising model as well as a three state opinion model could be used. Often it is required to modify the simple models according to real data in which case we have data-driven models. Recently, researchers have extended the Ising model using large-scale social media datasets to accurately describe how polarized online opinions form, evolve and eventually dissipate \cite{lu2021big}. In another context, the Ising model was applied directly to empirical data from scientific co-authorship networks, to study patterns of collaborative networks \cite{hurtado2021analysis}. This shows the model's versatility in diverse social contexts.

Future research may expand the scope of Ising-based sociophysics along multiple complementary directions. One promising avenue is to integrate these binary-agent models with modern machine learning techniques and large-scale social datasets, enabling data-driven calibration and improving their predictive realism. Another important extension is to move beyond binary states by considering multi-state or continuous-state agents (analogous to Potts models), which would capture more complex opinion spectra and decision options. In parallel, incorporating complex network topologies into Ising-like models can elucidate how social network structure influences collective outcomes, forging stronger links with network science. An important message that also emerges from recent studies is the robustness of the Ising-Glauber dynamics. Many modified update rules, whether in voter-like models or generalized interaction schemes, tend to converge back to Ising-like behavior, reinforcing the centrality of the Ising paradigm in modeling emergent social phenomena. Progress on these fronts will benefit from close interdisciplinary collaboration, bridging physics with sociology, psychology, and data science to ensure models remain empirically grounded. Finally, while our survey focused on selected domains, it is worth noting that Ising model 
has found its applications in many other contexts, for example, modeling  neural networks, mimicking the dynamics and collective animal behavior and decision-making, processes in biological systems including psychometry etc, which lie beyond the scope of the present article but offer rich opportunities for future exploration.\\


\noindent\textbf{Acknowledgments}

PS  thanks Soham Biswas, Soumyajyoti Biswas, Bikas K Chakrabarti, Arnab Chatterjee, Sudip Mukherjee, Parna Roy for collaborative works on related topics. The authors are also grateful to Amit Pradhan for help with figures.

\bibliography{sn-bibliography}

\end{document}